%% file: main_ICC_2020_-_Paper_II.tex
\documentclass[conference]{IEEEtran}
\usepackage[utf8]{inputenc}
\usepackage[english]{babel}
\usepackage[ruled, vlined]{algorithm2e}
\usepackage{algpseudocode}
\usepackage{amssymb}
\usepackage{array}
\usepackage{arydshln}
\usepackage{blkarray, bigstrut}
\usepackage{bm}
\usepackage{booktabs}
\usepackage{braket}
\usepackage{cite}
\usepackage{cuted}
\usepackage{enumitem}
\usepackage{graphicx}
\usepackage[mathscr]{euscript}
\usepackage[switch]{lineno}
\usepackage{mathtools}
\usepackage{multirow}
\usepackage{multicol}
\usepackage{mathtools}
\usepackage{xfrac}
\usepackage{framed} 
\usepackage[framed]{ntheorem}
\usepackage{pgfplots}
\usepackage{qtree}
\usepackage{adjustbox}
\usepackage{rotating}
\pgfplotsset{compat = 1.12}
\usepackage{stfloats}
\usepackage{subcaption}
\usepackage{url}
\usepackage{tikz}
\usepackage{tkz-berge}
\usepackage[framemethod=TikZ]{mdframed}
\usetikzlibrary{patterns}
\usetikzlibrary{spy}
\usetikzlibrary{shapes, backgrounds,calc, snakes}
\usetikzlibrary{decorations.pathreplacing}
\usetikzlibrary{matrix}
\usepackage{fancybox}
\usepackage[normalem]{ulem}
\usepgfplotslibrary{fillbetween}
\usetikzlibrary{shapes.geometric,shapes.arrows,decorations.pathmorphing}
\usetikzlibrary{matrix,chains,scopes,positioning,arrows,fit}

\newcommand{\argmin}{\arg\!\min}
\newcommand{\argmax}{\arg\!\max}

\definecolor{bblue}{rgb}{0.12392, 0.0490, 0.9588}
\definecolor{sskyblue}{rgb}{0.1529, 0.5882, 0.9216}
\definecolor{ggreen}{rgb}{0.7098, 0.95, 0.40781}
\definecolor{yyellow}{rgb}{0.9765, 0.9804, 0.0784}

\definecolor{color0}{HTML}{FF0147}
\definecolor{color1}{HTML}{F400DC}
\definecolor{color2}{HTML}{BA0DFF}
\definecolor{color3}{HTML}{5700E8}
\definecolor{color4}{HTML}{0B03FF}
\definecolor{color5}{HTML}{0957F4}
\definecolor{color6}{HTML}{03B3FF}
\definecolor{color7}{HTML}{08E8DA}
\definecolor{color8}{HTML}{07FF8E}
\definecolor{color9}{HTML}{51FF0A}

\definecolor{p1}{rgb}{1, 0.0667, 0}
\definecolor{p2}{rgb}{1, 0.24, 0}
\definecolor{p3}{rgb}{1, 0.349, 0}
\definecolor{p4}{rgb}{1, 0.490, 0}
\definecolor{p5}{rgb}{1, 0.631, 0}
\definecolor{p6}{rgb}{1, 0.792, 0}
\definecolor{p7}{rgb}{1, 0.933, 0}

\makeatletter
\def\newmaketag{%
	\def\maketag@@@##1{\hbox{\m@th\normalfont\normalsize##1}}%
}
\makeatother

\usepackage{todonotes}

\DeclareMathSizes{10pt}{8pt}{6pt}{5pt} 	

\makeatletter
\def\subsubsection{%
  \@startsection
    {subsubsection}                 
    {3}                             
    {\parskip}                    
    {3.5ex plus 1.5ex minus 1.5ex}  
    {0.7ex plus .5ex minus 0ex}     
    {\normalfont\normalsize\itshape}
}
\makeatother

\begin{document}
\title{\huge{Fairness-Aware Hybrid Precoding for mmWave NOMA Unicast/Multicast Transmissions in Industrial IoT}}

\author{
\IEEEauthorblockN{Luis F. Abanto-Leon, and Gek Hong (Allyson) Sim} 
\IEEEauthorblockA{Secure Mobile Networking (SEEMOO) Lab, Technische Universit\"{a}t Darmstadt, Germany} \{labanto, asim\}@seemoo.tu-darmstadt.de
}

\markboth{Journal of \LaTeX\ Class Files,~Vol.~14, No.~8, Jan~2019}%
{Shell \MakeLowercase{\textit{et al.}}: Bare Demo of IEEEtran.cls for IEEE Communications Society Journals}

\maketitle

\input{./text/abstract}


\IEEEpeerreviewmaketitle

\input{./text/introduction}

\input{./text/system_model}

\input{./text/result}

\input{./text/discussion}
\input{./text/conclusion}

\input{./text/acknowledgment}
\input{./text/appendix}

\ifCLASSOPTIONcaptionsoff
  \newpage
\fi

\bibliographystyle{IEEEtran}
\bibliography{ref}

\end{document}

%% file: text/abstract.tex
\begin{abstract}
This paper investigates dual-layer non-orthogonally superimposed transmissions for industrial internet of things (IoT) millimeter-wave communications. Essentially, the overlayer is a ubiquitous multicast signal devised to serve all the devices in coverage with a common message, i.e., critical control packet. The underlayer is a composite signal that consists of private unicast messages. Due to safety implications, it is critical that all devices can decode the multicast information. To ensure this requirement, we jointly optimize the \emph{hybrid precoder}, \emph{analog combiners}, \emph{power allocation}, and \emph{fairness}. Specifically, we incorporate a power splitting constraint between the two overlaid signals and enforce supplementary per-device constraints to guarantee multicast fairness. Performance is evaluated in terms of the spectral efficiency, multicast fairness, and bit error rate, thus corroborating the feasibility of our proposed scheme.
\end{abstract}

%% file: text/introduction.tex
\section{Introduction} \label{s1}

In factories, multiple industrial devices (e.g., sensors, actuators, programmable logic devices, robotic arms) are inherently hyper-connected via hard-wiring to ensure redundancy, safety and precise coordination among the different phases of a manufacturing process. Nevertheless, wired connections hinder extensive automation deployment and constrain the mechanics of mobile robotics. Considering the rapid densification of industrial devices, wired connections become less appealing for factories of the future (i.e., Industry 4.0). Thus, wireless information transmission is a viable alternative for these environments. However, guaranteeing high-performance in terms of fairness, spectral efficiency and reliability is a challenging task.



\subsection{Background and Motivation} \label{s1a}
Recent studies emphasize the importance of integrating non-orthogonal multiple access (NOMA) with the next-generation wireless technologies, e.g., massive multiple-input multiple-output (mMIMO) and millimeter-wave (mmWave) \cite{b1, b2}. NOMA can concurrently serve multiple devices within the same radio resource via non-orthogonal superposition of signals in power or code domain (e.g., \cite{b3, b4}), thus improving the spectral efficiency. By interweaving mMIMO, mmWave and NOMA, the expected next-generation networks throughput demands can be fulfilled while enabling simultaneous coexistence of heterogeneous connectivities. Owing to recent progress in mmWave technology, the mmWave spectrum is regarded as a plausible candidate to replace wires in industrial sectors. As a matter of fact, a measurement campaign reported that mmWave communications is feasible in industrial environments \cite{b5}. Furthermore, mmWave spectrum does not only provide high data rates (due to wide-band availability) but also characterizes for requiring antennas with a small footprint that can be easily embedded onto industrial devices and machinery. 

Due to high fabrication costs, hardware complexity and power consumption, mmWave fully-digital precoders are unaffordable. In contrast, more power-efficient hybrid (digital-analog) precoders have emerged as a functional solution, where a high-dimensional analog precoder (consisting of a low-cost phase shifts network) is connected in cascade with a low-dimensional digital precoder \cite{b6, b19}. Essentially, mmWave communications has become a tangible technology due to advancements in hybrid architectures. 

In this paper, we foresee an industrial scenario where a next generation Node B (gNodeB) transmits common multicast control beacons with superlative importance (e.g., critical safety packets, coordination messages) alongside less relevant private unicast information to each device (e.g., software or routine updates). To this aim, we propose the integration of NOMA, massive MIMO and mmWave. Specifically, NOMA is harnessed to transmit two overlaid signals (i.e., multicast and unicast) with different priorities. Further, mmWave provides wide-band spectrum that is efficiently exploited by massive MIMO. Thus, this integration can be leveraged to serve numerous industrial devices with a variety of data rate requirements while improving the spectral efficiency of the system. 


\subsection{Related Work} \label{s1b}
Power-domain NOMA (P-NOMA) is a comprehensive class of multiple-access technology capable of enhancing the spectrum utilization by means of superposing multiple signals with distinct power levels in the same time-frequency radio resource \cite{b8} (e.g., layered-division multiplexing (LDM) \cite{b7}, multi-user superposition transmission (MUST) \cite{b3}). Although promising, the deployment of P-NOMA in practical systems has been consistently neglected due to the implementation complexity for successive interference cancellation (SIC). Nevertheless, due to novel advancements in signal processing and silicon technology, P-NOMA has reemerged in digital terrestrial television (e.g., \cite{b7}) and wireless mobile communications (e.g., \cite{b3,b4}) domains as a feasible recourse to meet the ever-increasing data rate demands. For instance, \cite{b9} leverages LDM to transmit superposed broadcast/unicast signals in single-frequency networks. The authors of \cite{b10} investigate energy efficiency in MUST systems, where a base station with hybrid precoder conveys information to single-antenna receivers clustered in pairs. In \cite{b11}, dual-layer LDM broadcast/unicast transmissions with quality of service constraints (QoS) is researched, considering a system with fully-digital precoders and single-antenna receivers. The authors in \cite{b12} investigate the design of overlaid LDM unicast/multicast precoders with the aim of maximizing the sum-capacity in a scenario with a hybrid transmitter and multi-antenna users. In \cite{b13}, LDM broadcast/unicast transmissions with fully-digital transmitters and backhaul capacity constraints is investigated. In \cite{b14, b20}, a similar idea to that described in \cite{b11} is proposed, where two data layers are superposed. While the multicast and unicast data streams are originated independently in \cite{b11}, in \cite{b14, b20} the two data streams are interrelated. The unicast data for each user is split into common and private parts. Thus, one layer carries a common message consisting of the multicast information and the unicast common parts whereas the remaining layer conveys the unicast private messages only. Through this procedure, unicast inter-user interference (IUI) can be partially decoded and removed at each receiver (by exploiting the common unicast parts), thus further boosting the spectral efficiency. 

Although mentioned in a few prior works, transmit power splitting among unicast and multicast signals is not considered in the formulations. For instance, in \cite{b11}, explicit unicast and multicast QoS inequality constraints were imposed, thus removing the necessity of designing the power sharing between the signals. In \cite{b12}, this aspect was not considered, causing undesirable power allocation and information irrecoverability at the receivers. On the other hand, multicast fairness at each receiver has neither been researched in non-orthogonally overlaid multicast/unicast transmissions.

\subsection{Our Contributions} \label{s1c}

We study the spectral efficiency maximization problem in dual-layer LDM multicast/unicast systems, while considering hybrid precoders and analog combiners, subject to multicast fairness and transmit power constraints. To the best of our knowledge, \emph{we are the first to prioritize the multicast signal (over unicast) in dual-layer LDM multicast/unicast transmissions by means of a power-splitting mechanism while guaranteeing multicast fairness at each receiver}. These features are highly relevant for industrial IoT wireless networks to ensure successful decoding of the control signals. Specifically, we consider power splitting to allow the multicast signal to be received with higher power (than the aggregate unicast signal), thus ensuring proper operation of the SIC decoder. Also, to guarantee ubiquitous multicast service (i.e., delivery of critical control packets), we incorporate fairness constraints that guarantee decodability of the multicast information at each receiver, thus promoting reliability. Further, we assume a hybrid precoder at the transmitter and analog combiners at each receiver, where finite-resolution phase shifts are adopted. We propose two solutions for the problem described. Our first scheme, \texttt{PLDM-1}, designs independently the multicast precoder from the unicast precoders In the second approach, \texttt{PLDM-2}, the multicast precoder is obtained as a conic combination of the unicast precoding vectors. Thus, in \texttt{PLDM-1} each type of signal (multicast or unicast) is transmitted through a different precoder with distinct spatial radiation pattern. Contrastingly, \texttt{PLDM-2} repurposes the same precoder to multiplex the multicast and unicast signals with different powers. As a result, the two signals exhibit the the same spatial radiation pattern but with different power signatures.

Our paper is organized as follows. In Section \ref{s2}, we describe the system model whereas in Section \ref{s3}, we formulate the problem above mentioned. In Section \ref{s4}, our proposed solution is described in detail. Section \ref{s5} is devoted for simulation results. Section \ref{s6} discusses relevant implementation aspects while Section \ref{s7} summarizes our conclusions.

%% file: text/system_model.tex
\section{System Model} \label{s2}

\begin{figure}[!t]
	\centering
	\begin{tikzpicture}
	    \draw (0, 0) node[inner sep=0] {\includegraphics[width = 0.9\columnwidth]{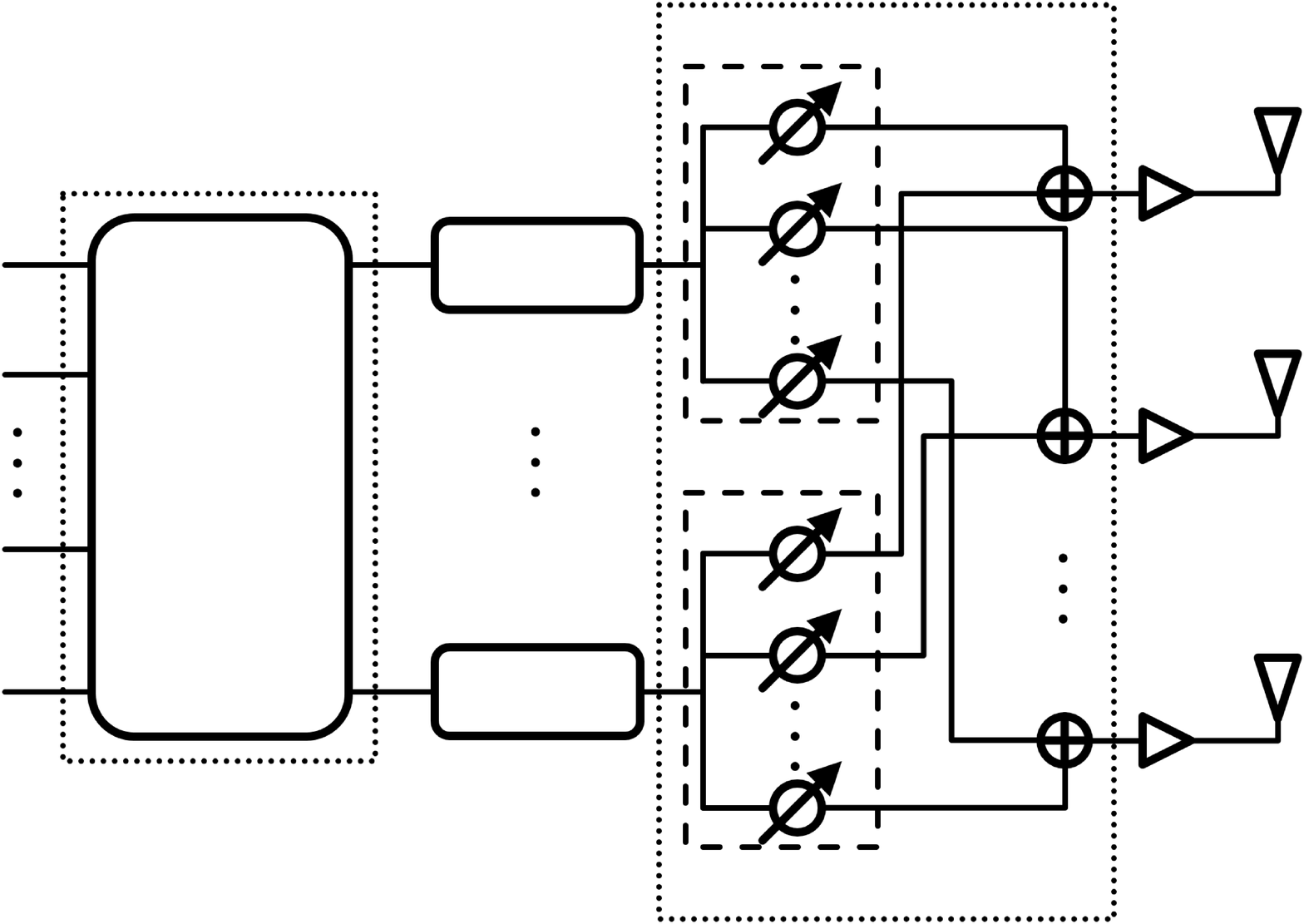}};
	    
	    \draw (1.4, 2.85) node{ \scriptsize{Analog precoder}};
	    \draw [decorate,decoration={brace,amplitude=4pt},xshift=-2pt,yshift=0pt] (2.9,-2.8) -- (-0.0,-2.8) node [black,midway,yshift=-9pt] {\footnotesize $ \mathbf{F} $};
	    
	    \draw (-2.6, 1.75) node{ \scriptsize{Digital precoder}};
	    \draw (-2.6, 0.3) node{ \scriptsize{Baseband}};
	    \draw (-2.6, 0.0) node{ \scriptsize{signal}};
	    \draw (-2.6, -0.3) node{ \scriptsize{processing}};
	    \draw [decorate,decoration={brace,amplitude=4pt},xshift=-2pt,yshift=0pt] (-1.5,-1.85) -- (-3.5,-1.85) node [black,midway,yshift=-9pt] {\footnotesize $ \left[ \mathbf{B} \vert \mathbf{m} \right] $};
	    
	    \draw (-0.7, 1.25) node{ \tiny{RF}};
	    \draw (-0.7, 1.1) node{ \tiny{chain $ {\scriptscriptstyle 1} $}};
	    
	    \draw (-0.7, -1.3) node{ \tiny{RF}};
	    \draw (-0.7, -1.45) node{ \tiny{chain $ {\scriptscriptstyle N^\mathrm{RF}_\mathrm{tx}} $}};
	    
	    \draw (0.8, 2.5) node{ $ {\scriptscriptstyle N_\mathrm{tx}} $ {\tiny shifters}};
	    
	    \draw (0.8, -2.55) node{ $ {\scriptscriptstyle N_\mathrm{tx}} $ {\tiny shifters}};
	    
	    \draw [decorate,decoration={brace,amplitude=4pt},yshift=0pt] (3.85,2.2) -- (3.85,-1.8) node [black,midway,xshift=12pt] {\footnotesize \rotatebox[]{-90}{$ N_\mathrm{tx} $ {\scriptsize antennas}}};
	    
	    \draw [decorate,decoration={brace,amplitude=4pt},yshift=0pt] (-3.9,-0.7) -- (-3.9,1.3) node [black,midway,xshift=-8pt] {\footnotesize {$ \mathbf{s} $}};
	    \draw (-4, -1.35) node{ $ z $};
	    
	\end{tikzpicture}
		\caption{Dual-layer unicast-multicast LDM system}
		\label{f1}	
\end{figure}

We assume a mmWave system, where a gNodeB serves $ K $ devices in the downlink. The transmitted signal consists of two non-orthogonal layers that overlap with different power levels\footnote{We only discuss herein \texttt{PLDM-1}. By redefining $ \mathbf{x} = \mathbf{F} \mathbf{B} \left[ \mathbf{I} \vert \mathbf{u} \right] \left[ \mathbf{s} \vert z \right]^T $, \texttt{PLDM-2} can be derived through a similar procedure. The elements of $ \mathbf{u} \in \mathbb{R}^{K \times 1} $ represent the weights to be optimized. Thus, $ \mathbf{m} = \mathbf{B} \mathbf{u} $.}. The \textit{primary layer} is a multicast signal that transports a shared high-priority packet intended for all the devices in coverage. The \emph{secondary layer} is a composite signal that contains multiple unicast messages, each intended for a distinct device. The gNodeB is equipped with a hybrid transmitter consisting of $ N_\mathrm{tx} $ transmit antennas and $ N^\mathrm{RF}_\mathrm{tx} \leq N_\mathrm{tx} $ radio frequency (RF) chains as shown in Fig. \ref{f1}. Essentially, a hybrid precoder is formed by a high-dimensional network of low-resolution constant-modulus phase shifters (analog precoder) that perform elementary beamsteering at the RF frequency. Interconnected in cascade, a low-dimensional mixed-signal constituent (digital precoder) operates at the baseband frequency performing more sophisticated beamforming \cite{b6, b22}. Each receiver possesses a single RF chain with $ N_\mathrm{rx} $ antennas, which enables analog receive beamforming. In addition, without loss of generality, we assume that $ N^\mathrm{RF}_\mathrm{tx} = K $. The downlink signal is represented by $ \mathbf{x} = \mathbf{F} \left[ \mathbf{B} \vert \mathbf{m} \right] \left[ \mathbf{s} \vert z \right]^T $, where $\mathbf{F} = \left[ \mathbf{f}_1,\mathbf{f}_2,\dots,\mathbf{f}_K \right] \in \mathbb{C}^{N_\mathrm{tx} \times K}$ is the analog precoder whereas $\mathbf{B} = \left[ \mathbf{b}_1,\mathbf{b}_2,\dots,\mathbf{b}_K \right] \in \mathbb{C}^{K \times K} $ and $ \mathbf{m} \in \mathbb{C}^{K \times 1}$ represent the unicast and multicast digital precoders, respectively. Also, $\mathbf{s} = \left[ s_1,s_2,\dots,s_K \right]^T \in \mathbb{C}^{K \times 1}$ denotes the unicast data symbols for the $ K $ devices and $z \in \mathbb{C}$ represents the common multicast symbol, with $\mathbb{E} \left\lbrace \mathbf{s} \mathbf{s}^H \right\rbrace = \mathbf{I} $ and $\mathbb{E} \left\lbrace z^{H} z \right\rbrace = 1$. Under the assumption of narrowband flat-fading, the signal received by the $k$-th device is given by
\begin{align} \label{e1}
	\begin{split}
		y_k = \underbrace{\mathbf{w}^H_k \mathbf{H}_k \mathbf{F} \mathbf{m} z}
		_{\text{common multicast signal}} &
		+ \underbrace{\mathbf{w}^H_k \mathbf{H}_k \mathbf{F} \mathbf{b}_k s_k}
		_{\text{unicast signal for device } k} 
		\\
		& + \underbrace{\mathbf{w}^H_k \mathbf{H}_k \mathbf{F} \sum_{j \neq k}{\mathbf{b}_j s_j}}
		_{\text{interference at device } k}		
		 + \underbrace{\mathbf{w}^H_k \mathbf{n}_k}_{\text{noise}}, 
	\end{split}
\end{align}
where $\mathbf{w}_k \in {\mathbb{C}}^{N_\mathrm{rx} \times 1}$ represents the analog combiner of the $k$-th device, $\mathbf{H}_k \in {\mathbb{C}}^{N_\mathrm{rx} \times N_\mathrm{tx}}$ denotes the downlink channel between the gNodeB and the $ k $-th device, whereas $\mathbf{n}_k \sim \mathcal{CN} \left( \mathbf{0}, {\sigma}^2 \mathbf{I} \right) $ denotes additive white Gaussian noise. At each receiver, the multicast symbol is decoded first by treating the aggregate unicast signals as interference. Subsequently, the multicast signal is reconstructed and then subtracted from $ y_k $ (by employing the decoded multicast symbol and the channel $ \mathbf{H}_k $). As a consequence, the remaining byproduct consists solely of unicast components and noise, from where each receiver $ k $ can decode its intended symbol $ s_k $. Thus, the signal--to--interference--plus--noise ratio (SINR) of the multicast and unicast signals are respectively defined as
\begin{equation} \label{e2}
	\tilde{\gamma}_k = \frac{\left| \mathbf{w}^H_k \mathbf{H}_k \mathbf{F} \mathbf{m} \right|^2 }
					   {
			                  \sum_j \left| \mathbf{w}^H_k \mathbf{H}_k \mathbf{F} {\mathbf{b}_j} \right|^2 + {\sigma}^2 \left\| \mathbf{w}_k \right\|^2_2 
		           	   }
\end{equation}
\begin{equation} \label{e3}
	\gamma_k = \frac{\left| \mathbf{w}^H_k \mathbf{H}_k \mathbf{F} {\mathbf{b}_k} \right|^2 }
			   {
					\sum_{j \neq k} \left| \mathbf{w}^H_k \mathbf{H}_k \mathbf{F} {\mathbf{b}_j} \right|^2 + {\sigma}^2 \left\| \mathbf{w}_k \right\|^2_2.
			   }
\end{equation}

\section{Problem Formulation} \label{s3}
Aiming at maximizing the aggregate multicast and unicast spectral efficiency, the optimization problem is formulated as
\begin{subequations} \label{e4}
	\begin{align}
		\mathcal{P}: & \max_{
				\substack{
							\left\lbrace \mathbf{w}_k \right\rbrace^K_{k = 1} , \\
							\left\lbrace \mathbf{f}_k \right\rbrace^K_{k = 1}, \\ 
							\left\lbrace \mathbf{b}_k \right\rbrace^K_{k = 1}, \mathbf{m}, \Delta
						 }
			   }
		& &
		\sum_k \log_2 \left( 1 + \tilde{\gamma}_k \right) + \log_2 \left( 1 + \gamma_k \right)  - C \Delta \label{e4a}
		\\
		\vspace{-0.2cm}
		& ~~~~~~ \mathrm{s.t.} & & \left| \tilde{\gamma}_k - {\gamma}_\mathrm{min} \right| \leq \Delta, \forall k \in \mathcal{K}, \label{e4b}
		\\
		& & & \tilde{\gamma}_1 \geq \tilde{\gamma}_2 \geq \dots \geq \tilde{\gamma}_K \geq \tilde{\gamma}_1, \label{e4c}
		\\
		& & & \left\| \mathbf{F} \mathbf{m} \right\|^2_2 /
			       \sum_k \left\| \mathbf{F} \mathbf{b}_k \right\|^2_2 \geq \beta, \label{e4d}
		\\
		& & & \left\| \mathbf{F} \mathbf{m} \right\|^2_2 + 
                           \sum_k \left\| \mathbf{F} \mathbf{b}_k \right\|^2_2 \leq P_\mathrm{tx}, \label{e4e}
		\\ 
		& & & \left[ \mathbf{F} \right]_{q,r} \in \mathcal{F}, q \in \mathcal{Q}, r \in \mathcal{R}, \label{e4f}
		\\ 
		& & & \left[ \mathbf{w}_k \right]_n \in \mathcal{W}, n \in \mathcal{N}, \forall k \in \mathcal{K}, \label{e4g}
		\\ 
		& & & \Delta \geq 0, \label{e4h}
	\end{align}
\end{subequations}
where (\ref{e4b}) confines the multicast SINRs to a narrow interval around a predefined threshold $ \gamma_\mathrm{min} $ (with deviation $ \Delta $), thus ensuring the control packet decoding. Constraint (\ref{e4c}) binds all the multicast SINRs together in order to promote fairness. Constraint (\ref{e4d}) splits the power among multicast and unicast signals (in the ratio of $ \beta \geq 1 $ to 1) with the purpose of enforcing higher priority on the multicast information and securing SIC decoding. Constraint (\ref{e4e}) restricts the transmit power to $ P_\mathrm{tx} $ while (\ref{e4f})-(\ref{e4g}) enforce the limitations of analog beamforming, i.e., only a small number of $ L_\mathrm{tx} $ (at the analog precoder) and $ L_\mathrm{rx} $ (at the analog combiners) constant-modulus phase shifts are allowed. Finally, (\ref{e4g}) enforces positiveness on $ \Delta $. Under sufficient power $ P_\mathrm{tx} $ (and large positive $ C $) then $ \Delta \rightarrow 0 $, since (\ref{e4b}) can be satisfied with equality. However, when $ P_\mathrm{tx} $ is insufficient, $ \Delta $ increases such that every $ \tilde{\gamma}_k $ is at most at $ \Delta $ from $ \gamma_\mathrm{min} $ while simultaneously enforcing fairness via (\ref{e4c}). Every element $ \left[ \mathbf{F} \right]_{q,r} $ of the analog precoder is constrained to a feasible set $ \mathcal{F} = \left\lbrace 1/{\sqrt{N_\mathrm{tx}}}, \dots, 1/{\sqrt{N_\mathrm{tx}}} e^{j \frac{2 \pi \left( L_\mathrm{tx} - 1 \right)}{L_\mathrm{tx}}} \right\rbrace $, where $ q \in \mathcal{Q} = \left\lbrace 1, \dots, N_\mathrm{tx} \right\rbrace $ and $ r \in \mathcal{R} = \left\lbrace 1, \dots, N^\mathrm{RF}_\mathrm{tx} \right\rbrace $. Similarly, every element $ \left[ \mathbf{w}_k \right]_n $ of the $ k $-th analog combiner is restricted to $ \mathcal{W} = \left\lbrace 1/{ \sqrt{N_\mathrm{rx}}}, \dots, 1/{\sqrt{N_\mathrm{rx}}} e^{j \frac{2 \pi \left( L_\mathrm{rx} - 1 \right)}{L_\mathrm{rx}}} \right\rbrace $, $ n \in \mathcal{N} = \left\lbrace 1, \dots, N_\mathrm{rx} \right\rbrace $.

\begin{mdframed}
\emph{Remark:} Note that (\ref{e4a}) aims to jointly maximize the sum-capacity of multicast and unicast signals. While maximization of the latter term is widely accepted, optimizing the multicast sum-capacity is non-standard. However, in our case the multicast term in (\ref{e4a}) is linked to (\ref{e4b})-(\ref{e4c}), thus enforcing the multicast SINRs to be equal and proximate to $ \gamma_\mathrm{min} $. Thus, the combined action of (\ref{e4a})-(\ref{e4c}) promotes multicast sum-capacity maximization and fairness improvement.
\end{mdframed}

\section{Proposed Solution} \label{s4}
We recast (\ref{e4}) as (\ref{e5}) without loss of optimality
\begin{subequations} \label{e5}
	\begin{align}
		\mathcal{P}_0: & \max_{
				\substack{
							\left\lbrace \mathbf{w}_k \right\rbrace^K_{k = 1},
							\left\lbrace \mathbf{f}_k \right\rbrace^K_{k = 1}, \\ 
							\left\lbrace p_k \right\rbrace^K_{k = 1},
							\left\lbrace \mathbf{v}_k \right\rbrace^K_{k = 1}, \mathbf{m}, \Delta
						 }
			   }
		& &
		\sum_k \tilde{\gamma}_k + \gamma_k - C \Delta \label{e5a}
		\\
		\vspace{-0.2cm}
		& ~~~~~~~~~~~ \mathrm{s.t.} & & \left| \tilde{\gamma}_k - \gamma_\mathrm{min} \right| \leq \Delta, \forall k \in \mathcal{K}, \label{e5b}
		\\
		& & & \tilde{\gamma}_1 \geq \tilde{\gamma}_2 \geq \dots \geq \tilde{\gamma}_K \geq \tilde{\gamma}_1, \label{e5c}
		\\
		& & & \left\| \mathbf{F} \mathbf{m} \right\|^2_2 /
			       \sum_k p_k \left\| \mathbf{F} \mathbf{v}_k \right\|^2_2 \geq \beta, \label{e5d}
		\\
		& & & \left\| \mathbf{F} \mathbf{m} \right\|^2_2 + 
                           \sum_k p_k \left\| \mathbf{F} \mathbf{v}_k \right\|^2_2 \leq P_\mathrm{tx}, \label{e5e}
		\\ 
		& & & \left[ \mathbf{F} \right]_{q,r} \in \mathcal{F}, q \in \mathcal{Q}, r \in \mathcal{R}, \label{e5f}
		\\ 
		& & & \left[ \mathbf{w}_k \right]_n \in \mathcal{W}, n \in \mathcal{N}, \forall k \in \mathcal{K}, \label{e5g}
		\\ 
		& & & \left\| \mathbf{v}_k \right\|^2_2 = 1, \forall k \in \mathcal{K}, \label{e5h}
		\\ 
		& & & p_k \geq 0, \forall k \in \mathcal{K}, \label{e5i}
		\\ 
		& & & \Delta \geq 0, \forall k \in \mathcal{K}, \label{e5j}
	\end{align}
\end{subequations}
where (\ref{e4a}) is transformed harnessing  $ \sum_k \log_2 \left( 1 + \gamma_k \right) \rightarrow \sum_k \gamma_k $ (refer to Appendix). Also, $ p_k $ is the power associated to the unit-power vector $ \mathbf{v}_k $, such that $ \mathbf{b}_k = \sqrt{p_k} \mathbf{v}_k $. Due to expressions involving multiplicative coupling (i.e., (\ref{e5a})-(\ref{e5e})) and non-convex domains (i.e., (\ref{e5f})-(\ref{e5g})), $ \mathcal{P}_0 $ is challenging to solve. Except for the  convex constraints (\ref{e5h})-(\ref{e5j}), (\ref{e5a})-(\ref{e5g}) are non-convex. In order to approach (\ref{e5}), we adopt a sequential optimization scheme that can only guarantee local optimality. Hence, $ \mathcal{P}_0 $ is decomposed into smaller sub-problems $ \mathcal{P}_1 $ and $ \mathcal{P}_2 $ that are independently and alternately optimized.

\subsection{Optimization of $ \mathbf{w}_k $, $ \mathbf{f}_k $ and $ \mathbf{v}_k $} \label{s4a}
We optimize $ \left\lbrace \mathbf{w}_k \right\rbrace^K_{k = 1} $, $ \left\lbrace \mathbf{f}_k \right\rbrace^K_{k = 1} $ and $ \left\lbrace \mathbf{v}_k \right\rbrace^K_{k = 1} $ to maximize the unicast sum-capacity while momentaneously disregarding the multicast constituent. Therefore, (\ref{e5b})-(\ref{e5c}) and (\ref{e5j}) are not considered in $ \mathcal{P}_1 $. Moreover, since $ \mathbf{m} $ and $ \left\lbrace p_k \right\rbrace^K_{k = 1} $ are optimized in $ \mathcal{P}_2 $, we can further suppress (\ref{e5d})-(\ref{e5e}) and (\ref{e5i})-(\ref{e5j}) since $ \mathbf{m} $ and $ \left\lbrace p_k \right\rbrace^K_{k = 1} $ can be finely tuned to satisfy such constraints in a latter stage. Thus, we define
\begin{subequations} \label{e6}
	\begin{align}
		\mathcal{P}_1: & \max_{
							\substack{
									\left\lbrace \mathbf{w}_k \right\rbrace^K_{k = 1}, \left\lbrace \mathbf{f}_k \right\rbrace^K_{k = 1}, \left\lbrace \mathbf{v}_k \right\rbrace^K_{k = 1}
						 			}
			   			      }
		& &
		\sum_k \gamma_k \label{e6a}
		\\
		\vspace{-0.2cm}
		& ~~~~~~~~~~~~~ \mathrm{s.t.} & & \left[ \mathbf{F} \right]_{q,r} \in \mathcal{F}, q \in \mathcal{Q}, r \in \mathcal{R}, \label{e6b}
		\\ 
		& & & \left[ \mathbf{w}_k \right]_n \in \mathcal{W}, n \in \mathcal{N}, \forall k \in \mathcal{K}, \label{e6c}
		\\ 
		& & & \left\| \mathbf{v}_k \right\|^2_2 = 1, \forall k \in \mathcal{K}. \label{e6d}
	\end{align}
\end{subequations}
Since (\ref{e6a}) entails coupling of parameters and (\ref{e6b})-(\ref{e6c}) are defined over non-convex sets, $ \mathcal{P}_1 $ is non-convex. On the other hand, the objective function $ \sum_k \gamma_k $ is a sum of fractional programs and therefore NP-complete. Although not guaranteeing optimality, a generally accepted practice is to express a sum of fractional programs in the substractive form \cite{b10}. Thus, we define the surrogate problem
\begin{subequations} \label{e7}
	\begin{align}
		\widetilde{\mathcal{P}}_1: & \max_{
									\substack{
											\left\lbrace \mathbf{w}_k \right\rbrace^K_{k = 1}, \left\lbrace \mathbf{f}_k \right\rbrace^K_{k = 1},  
											\left\lbrace \mathbf{v}_k \right\rbrace^K_{k = 1}
								 			}
					   			      }
		& &
		\sum_{k} p_k \left| \mathbf{w}^H_k \mathbf{H}_k \mathbf{F} {\mathbf{v}_k} \right|^2 \nonumber - \psi \sum_k \sigma^2 \\
		& & & - \psi \underbrace{\sum_{k} \sum_{j \neq k} p_j \left| \mathbf{w}^H_k \mathbf{H}_k \mathbf{F} {\mathbf{v}_j} \right|^2}_{\text{aggregate IUI of all $ K $ devices}} \label{e7a}
		\\
		\vspace{-0.2cm}
		& ~~~~~~~~~~~~~ \mathrm{s.t.} & & \left[ \mathbf{F} \right]_{q,r} \in \mathcal{F}, q \in \mathcal{Q}, r \in \mathcal{R}, \label{e7b}
		\\ 
		& & & \left[ \mathbf{w}_k \right]_n \in \mathcal{W}, n \in \mathcal{N}, \forall k \in \mathcal{K}, \label{e7c}
		\\ 
		& & & \left\| \mathbf{v}_k \right\|^2_2 = 1, \forall k \in \mathcal{K}, \label{e7d}
	\end{align}
\end{subequations}
which is obtained by subtracting the denominator from the numerator (with $ \left\| \mathbf{w}_k \right\|^2_2 = 1 $ due to (\ref{e6c})). We optimize $ \widetilde{\mathcal{P}}_1 $ by first maximizing $ \sum_{k} p_k \left| \mathbf{w}^H_k \mathbf{H}_k \mathbf{F} {\mathbf{v}_k} \right|^2 $ (in $ \widetilde{\mathcal{P}}_{1,1} $) and subsequently minimizing the aggregate IUI $ \sum_{k} \sum_{j \neq k} p_j \left| \mathbf{w}^H_k \mathbf{H}_k \mathbf{F} {\mathbf{v}_j} \right|^2 $ (in $ \widetilde{\mathcal{P}}_{1,2} $), in an alternate manner. For the same reasons mentioned above, $ \widetilde{\mathcal{P}}_1 $, $ \widetilde{\mathcal{P}}_{1,1} $ and $ \widetilde{\mathcal{P}}_{1,2} $ are also non-convex.
 
\subsubsection{Design of $\mathbf{w}_k$ and $\mathbf{f}_k$}
Disregarding the interference term in $ \widetilde{\mathcal{P}}_1 $, we maximize the aggregate power of the desired unicast signals at each receiver
\begin{subequations} \label{e8}
	\begin{align}
		\widetilde{\mathcal{P}}_{1,1}: & \max_{
											\left\lbrace \mathbf{w}_k \right\rbrace^K_{k = 1}, \left\lbrace \mathbf{f}_k \right\rbrace^K_{k = 1} 
					   			      		  }
		& &
		\sum_{k} p_k \left| \mathbf{w}^H_k \mathbf{H}_k \mathbf{F} {\mathbf{v}_k} \right|^2 \label{8a}
		\\
		\vspace{-0.2cm}
		& ~~~~~~~~ \mathrm{s.t.} & & \left[ \mathbf{F} \right]_{q,r} \in \mathcal{F}, q \in \mathcal{Q}, r \in \mathcal{R}, \label{e8b}
		\\ 
		& & & \left[ \mathbf{w}_k \right]_n \in \mathcal{W}, n \in \mathcal{N}, \forall k \in \mathcal{K}. \label{e8c}
	\end{align}
\end{subequations}

Without knowledge of $ \left\lbrace \mathbf{v}_k \right\rbrace^K_{k = 1} $, and since $ N^\mathrm{RF}_\mathrm{tx} \leq K $ we are in the position of maximizing the RF-to-RF channel gain $ \left| h_k \right|^2 = \left| \mathbf{w}^H_k \mathbf{H}_k \mathbf{f}_k \right|^2 $ for each device $ k $, where $ \mathbf{F} = \left[ \mathbf{f}_1,\mathbf{f}_2,\dots,\mathbf{f}_K \right] $. Thus, we define $ K $ sub-problems
\begin{subequations} \label{e9}
	\begin{align}
		\widetilde{\mathcal{P}}^{(k)}_{1,1}: & \max_{
												\mathbf{w}_k, \mathbf{f}_k 
					   			      		}
		& &
		\left| \mathbf{w}^H_k \mathbf{H}_k \mathbf{f}_k\right|^2 \label{9a}
		\\
		\vspace{-0.2cm}
		& ~
		~ \mathrm{s.t.} & & \left[ \mathbf{F} \right]_{q,r} \in \mathcal{F}, q \in \mathcal{Q}, r \in \mathcal{R}, \label{e9b}
		\\ 
		& & & \left[ \mathbf{w}_k \right]_n \in \mathcal{W}, n \in \mathcal{N}, \forall k \in \mathcal{K}, \label{e9c}
	\end{align}
\end{subequations}
that can be solved in parallel. The channel $\mathbf{H}_k$ is decomposed via singular value decomposition, i.e., $ \mathbf{H}_k = \left[ \mathbf{l}_k \vert \mathbf{L}_k \right] \boldsymbol{\Lambda}_k \left[ \mathbf{r}_k \vert \mathbf{R}_k \right]^H $, where $ \mathbf{l}_k $ and $ \mathbf{r}_k $ are the left and right singular vectors corresponding to the largest singular value \cite{b9} \cite{b12}. Then, $ \mathbf{w}_k $ and $ \mathbf{f}_k $ are selected such that $ \left[ \mathbf{w}_k \right]_n = \argmin_{\phi \in \mathcal{W}} \left| \phi - \left[ \mathbf{l}_k \right]_n \right|^2 = \argmax_{\phi \in \mathcal{W}} \mathfrak{Re} \left\lbrace \phi^* \left[ \mathbf{l}_k \right]_n \right\rbrace $ and $ \left[ \mathbf{f}_k \right]_l = \argmin_{\phi \in \mathcal{F}} \left| \phi - \left[ \mathbf{r}_k \right]_l \right|^2 = \argmax_{\phi \in \mathcal{F}} \mathfrak{Re} \left\lbrace \phi^* \left[ \mathbf{r}_k \right]_l \right\rbrace $, $ n \in \mathcal{N} $, $ q \in \mathcal{Q} $, $\forall k \in \mathcal{K} $, thus satisfying (\ref{e9b})-(\ref{e9c}). Essentially, $ \phi $ is chosen from $ \mathcal{W} $ or $ \mathcal{F} $, such that its phase is the closest to the phase of $ \left[ \mathbf{l}_k \right]_n $ or $ \left[ \mathbf{f}_k \right]_l $, respectively.

\subsubsection{Design of $\mathbf{v}_k$}
Upon suppressing the first term in $ \widetilde{\mathcal{P}}_1 $, we minimize the aggregate inter-user interference via
\begin{subequations} \label{e10}
	\begin{align}
		\widetilde{\mathcal{P}}_{1,2}: & \min_{
									\substack{
											\left\lbrace \mathbf{v}_k \right\rbrace^K_{k = 1}
								 			 }
					   			      	  }
		& &
		\sum_{k} \sum_{j \neq k} p_j \left| \mathbf{w}^H_k \mathbf{H}_k \mathbf{F} {\mathbf{v}_j} \right|^2 \label{e10a}
		\\
		\vspace{-0.2cm}
		& ~~~~ \mathrm{s.t.} & & \left\| \mathbf{v}_k \right\|^2_2 = 1, \forall k \in \mathcal{K}. \label{e10b}
	\end{align}
\end{subequations}

By harnessing zero-forcing (ZF) beamforming \cite{b11}, the unicast precoding vectors $ \mathbf{b}_k = p_k \mathbf{v}_k $ are obtained. As a result, the IUI can be removed to a great extent. To this purpose, we denote the effective baseband channel of device $ k $ as $ \mathbf{h}_k^{\mathrm{eff}} = \mathbf{w}_k^H \mathbf{H}_k \mathbf{F} $. Then, we obtain a set of unit-norm precoders $ \left\lbrace  \mathbf{v}_k \right\rbrace_{k=1}^K $ (by normalizing the ZF vectors $ \left\lbrace  \mathbf{b}_k \right\rbrace_{k=1}^K $), which minimize $ \sum_{k} \sum_{j \neq k} \left| \mathbf{h}_k^{\mathrm{eff}} {\mathbf{b}_j} \right|^2 = \sum_{k} \sum_{j \neq k} p_j \left| \mathbf{w}^H_k \mathbf{H}_k \mathbf{F} {\mathbf{v}_j} \right|^2 $ since $ \mathbf{h}^{\mathrm{eff}}_k \mathbf{b}_j \approx 0, \forall j \neq k $. For sufficiently large $ N_\mathrm{tx} $, the IUI term in (\ref{e10}) is negligible. Therefore, $ \tilde{\gamma}_k \approx \left| \mathbf{h}^{\mathrm{eff}}_k \mathbf{m} \right|^2 / \left( p_k \left| g_k \right|^2 + {\sigma}^2 \right)$ and $ \gamma_k \approx p_k \left| g_k \right|^2 / \sigma^2 $, where $ g_k = \mathbf{h}^{\mathrm{eff}}_k \mathbf{v}_k $ and $ \left\| \mathbf{w}_k \right\|^2_2 = 1 $.

\subsection{Optimization of $\mathbf{m}$ and $p_k$} \label{s4b}
We optimize the multicast precoder and unicast powers, 
\begin{subequations} \label{e11}
	\begin{align}
		\mathcal{P}_2: & \max_{
								\left\lbrace p_k \right\rbrace^K_{k = 1}, \mathbf{m}, \Delta
			   				  }
		& &
		\sum_k \tilde{\gamma}_k + \gamma_k - C \Delta \label{e11a}
		\\
		\vspace{-0.2cm}
		& ~~~~~~ \mathrm{s.t.} & & \left| \tilde{\gamma}_k - {\gamma}_\mathrm{min} \right| \leq \Delta, \forall k \in \mathcal{K}, \label{e11b}
		\\
		& & & \tilde{\gamma}_1 \geq \tilde{\gamma}_2 \geq \dots \geq \tilde{\gamma}_K \geq \tilde{\gamma}_1, \label{e11c}
		\\
		& & & \left\| \mathbf{F} \mathbf{m} \right\|^2_2 /
			       \sum_k p_k \left\| \mathbf{F} \mathbf{v}_k \right\|^2_2 \geq \beta, \label{e11d}
		\\
		& & & \left\| \mathbf{F} \mathbf{m} \right\|^2_2 + 
                           \sum_k p_k \left\| \mathbf{F} \mathbf{v}_k \right\|^2_2 \leq P_\mathrm{tx}, \label{e11e}
		\\ 
		& & & p_k \geq 0, \forall k \in \mathcal{K}, \label{e11f}
		\\
		& & & \Delta \geq 0. \label{e11g}
	\end{align}
\end{subequations}

The objective function (\ref{e11a}) is constructed as the sum of two quadratic-over-linear expressions, which are non-convex. Similarly, (\ref{e11b})-(\ref{e11d}) are of the same nature. On the other hand, (\ref{e11e})-(\ref{e11g}) are convex. To facilitate optimization, we introduce two sets of auxiliary parameters $ \left\lbrace \mu_k \right\rbrace^K_{k = 1} $, $ \left\lbrace \upsilon_k \right\rbrace^K_{k = 1} $ and define the following problem
\begin{subequations} \label{e12}
	\begin{align}
		\widetilde{\mathcal{P}}_2: & \max_{
										\substack{
											\left\lbrace p_k \right\rbrace^K_{k = 1},
											\left\lbrace \mu_k \right\rbrace^K_{k = 1}, \\ 
											\left\lbrace \upsilon_k \right\rbrace^K_{k = 1}, \mathbf{m}, \Delta
										 		}
							  } 
		& &
		\sum_k \mu_k + \upsilon_k - C \Delta \label{e12a}
		\\
		\vspace{-0.2cm}
		& ~~~~~~~~ \mathrm{s.t.} & & \left| \mathbf{h}^{\mathrm{eff}}_k \mathbf{m} \right|^2 / 
   		  \left( p_k \left| g_k \right|^2  + {\sigma}^2 \right) \geq \mu_k, \label{e12b}
		\\
		& & & p_k \left| g_k \right|^2 / \sigma^2 \geq \upsilon_k, \label{e12c}
		\\
		& & & \left\| \mathbf{F} \mathbf{m} \right\|^2_2 /
			\sum_{k} p_k \left\| \mathbf{F} \mathbf{v}_k \right\|^2_2 \geq \beta, \label{e12d}
		\\
		& & & \left\| \mathbf{F} \mathbf{m} \right\|^2_2 + 
		        \sum_{k} p_k \left\| \mathbf{F} \mathbf{v}_k \right\|^2_2 \leq P_\mathrm{tx}, \label{e12e}
		\\
		& & & \mu_1 \geq \mu_2 \geq \dots \geq \mu_K \geq \mu_1, \label{e12f}
		\\
		& & & \mu_k \leq \gamma_\mathrm{min} + \Delta, \label{e12g}
		\\
		& & & \mu_k \geq \gamma_\mathrm{min} - \Delta, \label{e12h}
		\\
		& & & \upsilon_k \geq 0, \label{e12i}
		\\
		& & & p_k \geq 0, \label{e12j}
		\\
		& & & \Delta \geq 0, \label{e12k}
	\end{align}
\end{subequations}
\begin{figure*}[b!]
	\hrulefill
	\begin{equation} \label{e13}
		2 \mathfrak{Re} \left\lbrace \left( p^{(t)}_k \left| g_k \right|^2 + {\sigma}^2 \left\| \mathbf{w}_k \right\|^2_2 \right) {\mathbf{m}^{(t)}}^H {\mathbf{h}^{\mathrm{eff}}_k}^H \mathbf{h}^{\mathrm{eff}}_k {\mathbf{m}}  \right\rbrace 
		- \left| \mathbf{h}^{\mathrm{eff}}_k {\mathbf{m}}^{(t)} \right|^2 \left( p_k \left| g_k \right|^2 + {\sigma}^2 \left\| \mathbf{w}_k \right\|^2_2 \right)
		- \mu_k \left( p^{(t)}_k \left| g_k \right|^2 + {\sigma}^2 \left\| \mathbf{w}_k \right\|^2_2 \right)^2 \geq 0.
	\end{equation}
	\begin{equation} \label{e14}
			2 \mathfrak{Re} \left\lbrace \left( \sum_{k} p^{(t)}_k \left\| \mathbf{F} \mathbf{v}_k \right\|^2_2 \right) {\mathbf{m}^{(t)}}^H \mathbf{F}^H \mathbf{F} \mathbf{m} \right\rbrace - \left(  \sum_{k} p_k \left\| \mathbf{F} \mathbf{v}_k \right\|^2_2 \right) {\mathbf{m}^{(t)}}^H \mathbf{F}^H \mathbf{F} {\mathbf{m}}^{(t)} - \beta\left( \sum_{k} p^{(t)}_k \left\| \mathbf{F} \mathbf{v}_k \right\|^2_2 \right) \geq 0.
		\end{equation}
\end{figure*}

The objective function (\ref{e12a}) defines the maximization of a linear function over $ \mu_k $ and $ \upsilon_k $, therefore it is convex. Constraints (\ref{e12c}) and (\ref{e12e})-(\ref{e12k}) are convex, whereas (\ref{e12b}) and (\ref{e12d}) are non-convex. In order to convexify $ \widetilde{\mathcal{P}}_2 $, we linearize the non-convex constraints (\ref{e12b}) and (\ref{e12d}) around a feasible point $ \left( \mathbf{m}^{(t)}, p^{(t)}_1, \cdots, p^{(t)}_K \right)  $. Thus, the convexified versions of (\ref{e12b}) and (\ref{e12d}) are shown in (\ref{e13}) and (\ref{e14}), respectively\footnote{When computing the gradients of real-valued expressions with respect to complex parameters (for linearization), we have employed the Wirtinger derivatives \cite{b21}.}. As a result, we optimize $ \widetilde{\mathcal{P}}^{(t)}_2 $ iteratively over a number of $ N_{\mathrm{iter}_2} $ iterations. In (\ref{e15}), we show the vectorized form of $ \widetilde{\mathcal{P}}_2 $ after linearization,
\begin{subequations} \label{e15}
	\begin{align}
		\widetilde{\mathcal{P}}^{(t)}_2: & \max_{\substack{
														\mathbf{m}, \mathbf{p}, \\
														\boldsymbol{\mu}, \boldsymbol{\upsilon}, \Delta
														  }
												}
		& &
		\mathbf{1}^T \boldsymbol{\mu} + \mathbf{1}^T \boldsymbol{\upsilon} - C \Delta \label{e15a}
  		\\
	  	\vspace{-0.2cm}
  		& ~~~ \mathrm{s.t.} & & 2 \mathfrak{Re} 
					\Big\{
						\mathrm{diag} \left( \mathbf{A} \mathbf{p}^{(t)} + \mathbf{d} \right)
						\left( \mathbf{I} \otimes  {{\mathbf{m}}^{(t)}}^H \right) \mathbf{C} \left( \mathbf{1} \otimes \mathbf{m} \right) \nonumber \Big\} \\
		& & & - \mathrm{diag} \left( \mathbf{A} \mathbf{p} + \mathbf{d} \right) \left( \mathbf{I} \otimes {\mathbf{m}^{(t)}}^H \right) \mathbf{C} \left( \mathbf{1} \otimes \mathbf{m}^{(t)} \right) -  \nonumber \\
		& & & \mathrm{diag} \left( \mathbf{A} \mathbf{p}^{(t)} + \mathbf{d} \right) \mathrm{diag} \left( \mathbf{A} \mathbf{p}^{(t)} + \mathbf{d} \right) \boldsymbol{\mu} \succcurlyeq \mathbf{0}, \label{e15b}
		\\
		& & & \left( \mathbf{A} \odot \left( \mathrm{diag} \left( \mathbf{d} \right) \right)^{-1} \right) \mathbf{p} \succcurlyeq \boldsymbol{\upsilon}, \label{e15c}
		\\
	 	& & & 2 \mathfrak{Re} 
		 	\left\lbrace 
		 	\left( \mathbf{c}^T \mathbf{p}^{(t)} {\mathbf{m}^{(t)}}^H \mathbf{F}^H 
		 	\mathbf{F} \mathbf{m} \right)
		 	\right\rbrace - \nonumber \\
		& & & \mathbf{c}^T \mathbf{p} {\mathbf{m}^{(t)}}^H \mathbf{F}^H \mathbf{F} {\mathbf{m}^{(t)}} - \left( \mathbf{c}^T \mathbf{p}^{(t)} \right)^2 \beta \geq 0, \label{e15d}
	 	\\
	 	& & & \left\| \mathbf{F} \mathbf{m} \right\|^2_2 + \sum_{k} p_k \left\| \mathbf{F} \mathbf{v}_k \right\|^2_2 \leq P_\mathrm{tx}, \label{e15e}
	 	\\
	 	& & & \left( \mathbf{I} - \widetilde{\mathbf{I}} \right) \boldsymbol{\mu} \succcurlyeq \mathbf{0}, \label{e15f}
	 	\\
		& & & \boldsymbol{\mu} \preccurlyeq \left( \gamma_\mathrm{min} + \Delta \right) \mathbf{1}, \label{e15g}
		\\
		& & & \boldsymbol{\mu} \succcurlyeq \left( \gamma_\mathrm{min} - \Delta \right) \mathbf{1}, \label{e15h}
		\\
		& & & \boldsymbol{\upsilon} \succcurlyeq \mathbf{0}, \label{e15i}
		\\
		& & & \mathbf{p} \succcurlyeq \mathbf{0}, \label{e15j}
		\\
		& & & \Delta \geq 0. \label{e15k}
	\end{align}
\end{subequations}
where $ \mathbf{A} = \mathrm{diag} \left( \left| g_1 \right|^2, \dots, \left| g_K \right|^2 \right) $, $ \mathbf{p} = \left[ p_1, \dots, p_K \right]^T $, $ \mathbf{d} = \sigma^2 \mathbf{1} $, $ \mathbf{C} = \mathrm{diag} \left( \left\| \mathbf{h}^{\mathrm{eff}}_1 \right\|^2_2, \dots, \left\| \mathbf{h}^{\mathrm{eff}}_K \right\|^2_2 \right) $, $ \mathbf{I} $ is the identity matrix, $ \widetilde{\mathbf{I}} $ is obtained by cyclically shifting all the columns of $ \mathbf{I} $ to the left only once, $ \mathbf{c} = \left[ \left\| \mathbf{F} \mathbf{v}_1 \right\|^2_2, \dots, \left\| \mathbf{F} \mathbf{v}_K \right\|^2_2 \right]^T$, $\boldsymbol{\mu} = \left[ \mu_1, \dots, \mu_K \right]^T $ and $ \boldsymbol{\upsilon} = \left[ \upsilon_1, \dots, \upsilon_K \right]^T $. Also, $ \otimes $ denotes the Kronecker product, whereas $ \odot $ represents component-wise multiplication. This formulation can be efficiently approached by convex optimization solvers. In our case, we use \texttt{CVX} and \texttt{SDPT3}.

%% file: text/result.tex
\section{Simulation Results} \label{s5}

\begin{figure*}[t!]
\begin{minipage}{0.33\textwidth}
	\begin{tikzpicture}
		\begin{axis}
			[
				xlabel = {$ P_\mathrm{tx}/{\sigma^2}$ [dB]},
				ylabel = {\small SE [bps/Hz]},
				xmin = -30,
				xmax = 5, 
				ymin = 0, 
				ymax = 37,
				ytick = {0, 5, 10, 15, 20, 25, 30, 35, 40},
				xtick = {-25, -10, 5},
				ticklabel style = {font=\scriptsize},
				width = 6.25cm,
				height = 7cm,
				legend columns = 1,
				legend pos = north east,
				legend style = {at = {(0.6,0.02)}, anchor = south west, font = \scriptsize, fill = none, align = left},
			]
			
			\addplot[color = black, mark = pentagon*, line width = 0.3pt, mark options = {scale = 0.8, fill = sskyblue, solid}, smooth] table {data/rateAverageSystemTotalOriginal.txt}; \addlegendentry{\texttt{PLDM-0}}

			\addplot[color = black, mark = triangle*, line width = 0.3pt, mark options = {scale = 0.8, fill = color1, solid}, smooth] table {data/rateAverageSystemTotalProposed1.txt}; \addlegendentry{\texttt{PLDM-1}}
			
			\addplot[color = black, mark = *, line width = 0.3pt, mark options = {scale = 0.5, fill = yellow, solid}, smooth] table {data/rateAverageSystemTotalProposed2.txt}; \addlegendentry{\texttt{PLDM-2}}

		\end{axis}

		\begin{axis}
			[
				xmin = -31,
				xmax = -29,
				ymin = 0.10,
				ymax = 0.30,
				width = 2.2cm,
				height = 3.0cm,
				ytick = {0.10, 0.15, 0.20, 0.25, 0.30},
				yticklabels = {0.10, 0.15, 0.20, 0.25, 0.30},
				xtick = {-30},
				xticklabels = {-30},
				tick label style={font=\fontsize{5}{4}\selectfont,},
				shift={(0.7cm,3.8cm)},axis background/.style={fill=white}
			]
			
			\addplot[color = black, mark = pentagon*, line width = 0.3pt, mark options = {scale = 0.8, fill = sskyblue, solid}, smooth] table {data/rateAverageSystemTotalOriginal.txt}; 
			
			\addplot[color = black, mark = triangle*, line width = 0.3pt, mark options = {scale = 0.8, fill = color1, solid}, smooth] table {data/rateAverageSystemTotalProposed1.txt}; 
			
			\addplot[color = black, mark = *, line width = 0.3pt, mark options = {scale = 0.5, fill = yellow, solid}, smooth] table {data/rateAverageSystemTotalProposed2.txt}; 
		\end{axis}
		
		\begin{axis}
			[
				xmin = -26,
				xmax = -24,
				ymin = 0.40,
				ymax = 0.80,
				width = 2.2cm,
				height = 3.0cm,
				ytick = {0.40, 0.50, 0.60, 0.70, 0.80},
				yticklabels = {0.40, 0.50, 0.60, 0.70, 0.80},
				xtick = {-25},
				xticklabels = {-25},
				tick label style={font=\fontsize{5}{4}\selectfont,},
				shift={(1.9cm,3.8cm)},axis background/.style={fill=white}
			]
			
			\addplot[color = black, mark = pentagon*, line width = 0.3pt, mark options = {scale = 0.8, fill = sskyblue, solid}, smooth] table {data/rateAverageSystemTotalOriginal.txt}; 
			
			\addplot[color = black, mark = triangle*, line width = 0.3pt, mark options = {scale = 0.8, fill = color1, solid}, smooth] table {data/rateAverageSystemTotalProposed1.txt}; 
			
			\addplot[color = black, mark = *, line width = 0.3pt, mark options = {scale = 0.5, fill = yellow, solid}, smooth] table {data/rateAverageSystemTotalProposed2.txt}; 
		\end{axis}
		
		\begin{axis}
			[
				xmin = -21,
				xmax = -19,
				ymin = 1.25,
				ymax = 2.05,
				width = 2.2cm,
				height = 3.0cm,
				ytick = {1.25, 1.45, 1.65, 1.85, 2.05},
				yticklabels = {1.25, 1.45, 1.65, 1.85, 2.05},
				xtick = {-20},
				xticklabels = {-20},
				tick label style={font=\fontsize{5}{4}\selectfont,},
				shift={(0.7cm,1.65cm)},axis background/.style={fill=white}
			]
			
			\addplot[color = black, mark = pentagon*, line width = 0.3pt, mark options = {scale = 0.8, fill = sskyblue, solid}, smooth] table {data/rateAverageSystemTotalOriginal.txt}; 
			
			\addplot[color = black, mark = triangle*, line width = 0.3pt, mark options = {scale = 0.8, fill = color1, solid}, smooth] table {data/rateAverageSystemTotalProposed1.txt}; 
			
			\addplot[color = black, mark = *, line width = 0.3pt, mark options = {scale = 0.5, fill = yellow, solid}, smooth] table {data/rateAverageSystemTotalProposed2.txt}; 
		\end{axis}
		
		\begin{axis}
			[
				xmin = -16,
				xmax = -14,
				ymin = 3.2,
				ymax = 5.2,
				width = 2.2cm,
				height = 3.0cm,
				ytick = {3.20, 3.70, 4.20, 4.70, 5.20},
				yticklabels = {3.20, 3.70, 4.20, 4.70, 5.20},
				xtick = {-15},
				xticklabels = {-15},
				tick label style={font=\fontsize{5}{4}\selectfont,},
				shift={(1.9cm,1.65cm)},axis background/.style={fill=white}
			]
			
			\addplot[color = black, mark = pentagon*, line width = 0.3pt, mark options = {scale = 0.8, fill = sskyblue, solid}, smooth] table {data/rateAverageSystemTotalOriginal.txt}; 
			
			\addplot[color = black, mark = triangle*, line width = 0.3pt, mark options = {scale = 0.8, fill = color1, solid}, smooth] table {data/rateAverageSystemTotalProposed1.txt}; 
			
			\addplot[color = black, mark = *, line width = 0.3pt, mark options = {scale = 0.5, fill = yellow, solid}, smooth] table {data/rateAverageSystemTotalProposed2.txt}; 
		\end{axis}
		
	\end{tikzpicture}
	\caption{Overall SE of the system}
	\label{f2}
\end{minipage}
\begin{minipage}{0.33\textwidth}
	\begin{tikzpicture}
		\begin{axis}
			[
				xlabel = {$ P_\mathrm{tx}/{\sigma^2}$ [dB]},
				ylabel = {\small SE [bps/Hz]},
				xmin = -30,
				xmax = 10, 
				ymin = 0, 
				ymax = 37,
				ytick = {0, 5, 10, 15, 20, 25, 30, 35, 40},
				xtick = {-30, -25, -20, -15, -10, -5, 0, 5, 10},
				ticklabel style = {font=\scriptsize},
				width = 6.25cm,
				height = 7cm,
				legend columns = 1,
				legend pos = north east,
				legend style = {at = {(0.7,0.01)}, anchor = south west, font = \scriptsize, fill = white, align = left},
			]
			
			\addlegendimage{color = black, mark = pentagon*, mark options = {scale = 0.8, fill = sskyblue, solid}, line width = 0.3pt, smooth, only marks} \addlegendentry{\texttt{PLDM-0}}
			\addlegendimage{color = black, mark = triangle*, mark options = {scale = 0.8, fill = color1, solid}, line width = 0.3pt, smooth, only marks}	\addlegendentry{\texttt{PLDM-1}}
			\addlegendimage{color = black, mark = *, mark options = {scale = 0.5, fill = yellow, solid}, line width = 0.3pt, smooth, only marks} \addlegendentry{\texttt{PLDM-2}}
			
			\addplot[color = black, mark = pentagon*, mark options = {scale = 0.8, fill = sskyblue, solid}, line width = 0.3pt, smooth, densely dotted] table {data/rateAverageSystemUnicastOriginal.txt}; 
			
			\addplot[color = black, mark = pentagon*, mark options = {scale = 0.8, fill = sskyblue, solid}, line width = 0.3pt, smooth, dashed] table {data/rateAverageSystemMulticastOriginal.txt}; 
			
			\addplot[color = black, mark = triangle*, mark options = {scale = 0.8, fill = color1, solid}, line width = 0.3pt, smooth, densely dotted] table {data/rateAverageSystemUnicastProposed1.txt}; 
			
			\addplot[color = black, mark = triangle*, mark options = {scale = 0.8, fill = color1, solid}, line width = 0.3pt, smooth, dashed] table {data/rateAverageSystemMulticastProposed1.txt}; 
			
			\addplot[color = black, mark = *, mark options = {scale = 0.5, fill = yellow, solid}, line width = 0.3pt, smooth, densely dotted] table {data/rateAverageSystemUnicastProposed2.txt}; 
			
			\addplot[color = black, mark = *, mark options = {scale = 0.5, fill = yellow, solid}, line width = 0.3pt, smooth, dashed] table {data/rateAverageSystemMulticastProposed2.txt}; 
			
			\addplot[color = red, mark=none, style = dashed, smooth] coordinates {(-30,12.3443) (25,12.3443)};
			
			\node[above,red] at (-18.5,11.75) {\scriptsize aggregate multicast SE};
			
		\end{axis}
		
		\begin{axis}
			[
				xmin = -25.5,
				xmax = -24.5,
				ymin = 0,
				ymax = 0.8,
				width = 2.1cm,
				height = 3.6cm,
				ytick = {0.0,0.2,0.4,0.6,0.8},
				yticklabels = {0.0,0.2,0.4,0.6,0.8},
				xtick = {-25, -20 ,-15},
				xticklabels = {-25, -20 ,-15},
				tick label style={font=\fontsize{5}{4}\selectfont,},
				shift={(0.4cm,2.75cm)},axis background/.style={fill=white}
			]
			
			\addplot[color = black, mark = pentagon*, mark options = {scale = 0.8, fill = sskyblue, solid}, line width = 0.3pt, smooth, densely dotted] table {data/rateAverageSystemUnicastOriginal.txt}; 
			
			\addplot[color = black, mark = pentagon*, mark options = {scale = 0.8, fill = sskyblue, solid}, line width = 0.3pt, smooth, dashed] table {data/rateAverageSystemMulticastOriginal.txt}; 
			
			\addplot[color = black, mark = triangle*, mark options = {scale = 0.8, fill = color1, solid}, line width = 0.3pt, smooth, densely dotted] table {data/rateAverageSystemUnicastProposed1.txt}; 
			
			\addplot[color = black, mark = triangle*, mark options = {scale = 0.8, fill = color1, solid}, line width = 0.3pt, smooth, dashed] table {data/rateAverageSystemMulticastProposed1.txt}; 
			
			\addplot[color = black, mark = *, mark options = {scale = 0.5, fill = yellow, solid}, line width = 0.3pt, smooth, densely dotted] table {data/rateAverageSystemUnicastProposed2.txt}; 
			
			\addplot[color = black, mark = *, mark options = {scale = 0.5, fill = yellow, solid}, line width = 0.3pt, smooth, dashed] table {data/rateAverageSystemMulticastProposed2.txt}; 

		\end{axis}
		
		\begin{axis}
			[
				xmin = -20.5,
				xmax = -19.5,
				ymin = 0,
				ymax = 2,
				width = 2.1cm,
				height = 3.6cm,
				ytick = {0.0,0.5,1.0,1.5,2.0},
				yticklabels = {0.0,0.5,1.0,1.5,2.0},
				xtick = {-25, -20 ,-15},
				xticklabels = {-25, -20 ,-15},
				tick label style={font=\fontsize{5}{4}\selectfont,},
				shift={(1.4cm,2.75cm)},axis background/.style={fill=white}
			]
			
			\addplot[color = black, mark = pentagon*, mark options = {scale = 0.8, fill = sskyblue, solid}, line width = 0.3pt, smooth, densely dotted] table {data/rateAverageSystemUnicastOriginal.txt}; 
			
			\addplot[color = black, mark = pentagon*, mark options = {scale = 0.8, fill = sskyblue, solid}, line width = 0.3pt, smooth, dashed] table {data/rateAverageSystemMulticastOriginal.txt}; 
			
			\addplot[color = black, mark = triangle*, mark options = {scale = 0.8, fill = color1, solid}, line width = 0.3pt, smooth, densely dotted] table {data/rateAverageSystemUnicastProposed1.txt}; 
			
			\addplot[color = black, mark = triangle*, mark options = {scale = 0.8, fill = color1, solid}, line width = 0.3pt, smooth, dashed] table {data/rateAverageSystemMulticastProposed1.txt}; 
			
			\addplot[color = black, mark = *, mark options = {scale = 0.5, fill = yellow, solid}, line width = 0.3pt, smooth, densely dotted] table {data/rateAverageSystemUnicastProposed2.txt}; 
			
			\addplot[color = black, mark = *, mark options = {scale = 0.5, fill = yellow, solid}, line width = 0.3pt, smooth, dashed] table {data/rateAverageSystemMulticastProposed2.txt}; 
		
		\end{axis}
		
		\begin{axis}
			[
				xmin = -15.5,
				xmax = -14.5,
				ymin = 0.2,
				ymax = 5,
				width = 2.1cm,
				height = 3.6cm,
				ytick = {0.2,1.4,2.6,3.8,5.0},
				yticklabels = {0.2,1.4,2.6,3.8,5.0},
				xtick = {-25, -20 ,-15},
				xticklabels = {-25, -20 ,-15},
				tick label style={font=\fontsize{5}{4}\selectfont,},
				shift={(2.4cm,2.75cm)},axis background/.style={fill=white}
			]
			
			\addplot[color = black, mark = pentagon*, mark options = {scale = 0.8, fill = sskyblue, solid}, line width = 0.3pt, smooth, densely dotted] table {data/rateAverageSystemUnicastOriginal.txt}; 
			
			\addplot[color = black, mark = pentagon*, mark options = {scale = 0.8, fill = sskyblue, solid}, line width = 0.3pt, smooth, dashed] table {data/rateAverageSystemMulticastOriginal.txt}; 
			
			\addplot[color = black, mark = triangle*, mark options = {scale = 0.8, fill = color1, solid}, line width = 0.3pt, smooth, densely dotted] table {data/rateAverageSystemUnicastProposed1.txt}; 
			
			\addplot[color = black, mark = triangle*, mark options = {scale = 0.8, fill = color1, solid}, line width = 0.3pt, smooth, dashed] table {data/rateAverageSystemMulticastProposed1.txt}; 
			
			\addplot[color = black, mark = *, mark options = {scale = 0.5, fill = yellow, solid}, line width = 0.3pt, smooth, densely dotted] table {data/rateAverageSystemUnicastProposed2.txt}; 
			
			\addplot[color = black, mark = *, mark options = {scale = 0.5, fill = yellow, solid}, line width = 0.3pt, smooth, dashed] table {data/rateAverageSystemMulticastProposed2.txt}; 
		
		\end{axis}

		\begin{axis}
			[
				xmin = -9.5,
				xmax = -10.5,
				ymin = 0.2,
				ymax = 10,
				width = 2.1cm,
				height = 3.6cm,
				ytick = {0.2, 2.7, 5.2, 7.7, 10},
				yticklabels = {0.2, 2.7, 5.2, 7.7, 10},
				xtick = {-15, -10 ,-5},
				xticklabels = {-15, -10 ,-5},
				tick label style={font=\tiny},
				shift={(3.4cm,2.75cm)},axis background/.style={fill=white},
				legend columns = 3,
				legend pos = north east,
				legend style = {at = {(-4.3,1.05)}, anchor = south west, font = \tiny, text depth = .ex, fill = white, align = left},
			]
			
			\addlegendimage{dashed}; \addlegendentry{Multicast}
			\addlegendimage{densely dotted}; \addlegendentry{Unicast}
			
			\addplot[color = black, mark = pentagon*, mark options = {scale = 0.8, fill = sskyblue, solid}, line width = 0.3pt, smooth, densely dotted] table {data/rateAverageSystemUnicastOriginal.txt}; 
			
			\addplot[color = black, mark = pentagon*, mark options = {scale = 0.8, fill = sskyblue, solid}, line width = 0.3pt, smooth, dashed] table {data/rateAverageSystemMulticastOriginal.txt}; 
			
			\addplot[color = black, mark = triangle*, mark options = {scale = 0.8, fill = color1, solid}, line width = 0.3pt, smooth, densely dotted] table {data/rateAverageSystemUnicastProposed1.txt}; 
			
			\addplot[color = black, mark = triangle*, mark options = {scale = 0.8, fill = color1, solid}, line width = 0.3pt, smooth, dashed] table {data/rateAverageSystemMulticastProposed1.txt}; 
			
			\addplot[color = black, mark = *, mark options = {scale = 0.5, fill = yellow, solid}, line width = 0.3pt, smooth, densely dotted] table {data/rateAverageSystemUnicastProposed2.txt}; 
			
			\addplot[color = black, mark = *, mark options = {scale = 0.5, fill = yellow, solid}, line width = 0.3pt, smooth, dashed] table {data/rateAverageSystemMulticastProposed2.txt}; 

		\end{axis}
	\end{tikzpicture}
	\caption{Disaggregated SE of the system}
	\label{f3}
\end{minipage}
\begin{minipage}{0.33\textwidth}
	\begin{tikzpicture}
		\begin{axis}
			[
				xlabel = {$ P_\mathrm{tx}/{\sigma^2} $ [dB]},
				ylabel = {\small SE [bps/Hz]},
				xmin = -30,
				xmax = 10, 
				ymin = 0, 
				ymax = 3.2,
				ytick = {0.0, 0.5, 1.0, 1.5, 2.0, 2.5, 3.0},
				xtick = {-30, -25, -20, -15, -10, -5, 0, 5, 10},
				ticklabel style = {font=\scriptsize},
				width = 6.25cm,
				height = 7cm, 
				legend columns = 1,
				legend pos = north east,
				legend style = {at = {(0.55,0.02)}, anchor = south west, font = \scriptsize, text width = 0.8cm, text height = 0.03cm, text depth = .ex, fill = none, align = left},
			]
		
			\addlegendimage{color = black, mark = pentagon*, mark options = {scale = 0.8, fill = sskyblue, solid}, line width = 0.3pt, smooth} \addlegendentry{\texttt{PLDM-0}}
			
			\addlegendimage{color = black, mark = triangle*, mark options = {scale = 0.8, fill = color1, solid}, line width = 0.3pt, smooth} \addlegendentry{\texttt{PLDM-1}}
			
			\addlegendimage{color = black, mark = *, mark options = {scale = 0.5, fill = yellow, solid}, line width = 0.3pt, smooth} \addlegendentry{\texttt{PLDM-2}}
			
			\addplot[color = black, mark = pentagon*, mark options = {scale = 0.8, fill = sskyblue, solid}, line width = 0.3pt, smooth, dashed] table {data/rateAverageUserMulticastOriginal.txt}; 
			
			\addplot[name path = PoUpper, color = sskyblue!40, line width = 0pt, smooth] table {data/rateAveragePlusStdDevUserMulticastOriginal.txt}; 

			\addplot[name path = PoLower, color = sskyblue!40, line width = 0pt, smooth] table {data/rateAverageMinusStdDevUserMulticastOriginal.txt}; 
			
			\addplot[sskyblue!40, fill opacity = 0.4, smooth] fill between[of = PoLower and PoUpper];
			
			\addplot[color = black, mark = triangle*, mark options = {scale = 0.8, fill = color1, solid}, line width = 0.3pt, smooth, dashed] table {data/rateAverageUserMulticastProposed1.txt}; 
			
			\addplot[name path = P1Upper, color = color1!40, line width = 0pt, smooth] table {data/rateAveragePlusStdDevUserMulticastProposed1.txt}; 

			\addplot[name path = P1Lower, color = color1!40, line width = 0pt, smooth] table {data/rateAverageMinusStdDevUserMulticastProposed1.txt}; 
			
			\addplot[color1!40, fill opacity = 0.4, smooth] fill between[of = P1Lower and P1Upper];
			
			\addplot[color = black, mark = *, mark options = {scale = 0.5, fill = yellow, solid}, line width = 0.3pt, smooth, dashed] table {data/rateAverageUserMulticastProposed2.txt}; 
			
			\addplot[name path = P2Upper, color = yellow!40, line width = 0pt, smooth] table {data/rateAveragePlusStdDevUserMulticastProposed2.txt}; 
			
			\addplot[name path = P2Lower, color = yellow!40, line width = 0pt, smooth] table {data/rateAverageMinusStdDevUserMulticastProposed2.txt}; 
			
			\addplot[yellow!40, fill opacity = 0.4, smooth] fill between[of = P2Lower and P2Upper];
			
			\addplot[color = red, mark=none, style = densely dotted, smooth] coordinates {(-30,2) (25,2)};
			
			\node[above,red] at (-18.8,1.95) {\scriptsize multicast target per device};
	
		\end{axis}
	\end{tikzpicture}
	\caption{Multicast SE per device}
	\label{f4}
\end{minipage}	
\end{figure*}
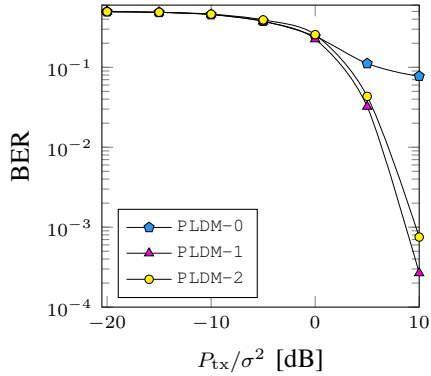
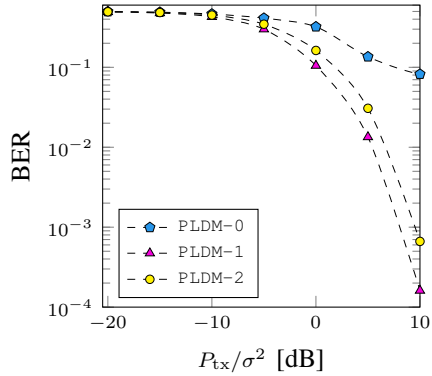
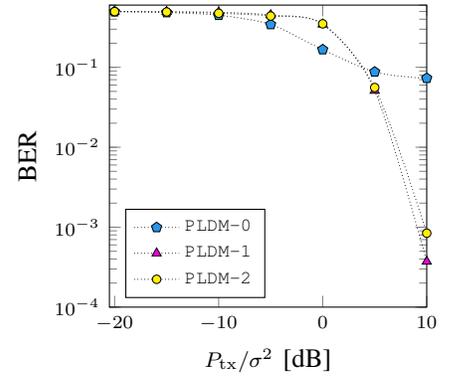
\begin{figure*}[th!]
	\centering
	\begin{subfigure}{0.32\textwidth}
		\centering
		\begin{tikzpicture}
		\begin{semilogyaxis}
		[
		xlabel = {$ P_\mathrm{tx}/{\sigma^2}$ [dB]},
		ylabel = {BER},
		ymode = log,
		xmin = -20.5,
		xmax = 10, 
		ymin = 1E-4, 
		ymax = 0.6,
		width = 5.8cm,
		height = 5.6cm,
		ticklabel style = {font=\scriptsize},
		legend columns = 1,
		legend pos = north east,
		legend style = {at = {(0.05,0.03)}, anchor = south west, font = \scriptsize, fill = white, align = left},
		]

		
		
		\addplot[color = black, mark = pentagon*, mark options = {scale = 1, fill = sskyblue, solid}, line width = 0.3pt, smooth] table {data/berAverageSystemTotalOriginal.txt};
		\addlegendentry[]{\texttt{PLDM-0}} 
		
		\addplot[color = black, mark = triangle*, mark options = {scale = 1, fill = color1, solid}, line width = 0.3pt, smooth] table {data/berAverageSystemTotalProposed1.txt}; 
		\addlegendentry[]{\texttt{PLDM-1}} 
		
		\addplot[color = black, mark = *, mark options = {scale = 0.8, fill = yellow, solid}, line width = 0.3pt, smooth] table {data/berAverageSystemTotalProposed2.txt}; 
		\addlegendentry[]{\texttt{PLDM-2}} 
		
		\end{semilogyaxis}
		\end{tikzpicture}
		\centering
		\caption{Aggregate}
	\end{subfigure}%
	\hspace{2mm}
	\begin{subfigure}{0.32\textwidth}
		\centering
		\begin{tikzpicture}
		\begin{semilogyaxis}
		[
		xlabel = {$ P_\mathrm{tx}/{\sigma^2}$ [dB]},
		ylabel = {BER},
		xmin = -20.5,
		xmax = 10, 
		ymin = 1E-4, 
		ymax = 0.6,
		width = 5.8cm,
		height = 5.6cm,
		ticklabel style = {font=\scriptsize},
		legend columns = 1,
		legend pos = north east,
		legend style = {at = {(0.05,0.03)}, anchor = south west, font = \scriptsize, fill = white, align = left},
		]

		\addplot[color = black, mark = pentagon*, mark options = {scale = 1, fill = sskyblue, solid}, line width = 0.3pt, smooth, dashed] table {data/berAverageSystemMulticastOriginal.txt}; 
		\addlegendentry[]{\texttt{PLDM-0}} 
		
		\addplot[color = black, mark = triangle*, mark options = {scale = 1, fill = color1, solid}, line width = 0.3pt, smooth, dashed] table {data/berAverageSystemMulticastProposed1.txt}; 
		\addlegendentry[]{\texttt{PLDM-1}} 
		
		\addplot[color = black, mark = *, mark options = {scale = 0.8, fill = yellow, solid}, line width = 0.3pt, smooth, dashed] table {data/berAverageSystemMulticastProposed2.txt}; 
		\addlegendentry[]{\texttt{PLDM-2}} 
		
		\end{semilogyaxis}
		\end{tikzpicture}
		\centering
		\caption{Multicast}
	\end{subfigure}%
	\hspace{2mm}
	\begin{subfigure}{0.32\textwidth}
		\centering
		\begin{tikzpicture}
		\begin{semilogyaxis}
		[
		xlabel = {$ P_\mathrm{tx}/{\sigma^2}$ [dB]},
		ylabel = {BER},
		xmin = -20.5,
		xmax = 10, 
		ymin = 1E-4, 
		ymax = 0.6,
		width = 5.8cm,
		height = 5.6cm,
		ticklabel style = {font=\scriptsize},
		legend columns = 1,
		legend pos = north east,
		legend style = {at = {(0.05,0.03)}, anchor = south west, font = \scriptsize, fill = white, align = left},
		]

		\addplot[color = black, mark = pentagon*, mark options = {scale = 1, fill = sskyblue, solid}, line width = 0.3pt, smooth, densely dotted] table {data/berAverageSystemUnicastOriginal.txt}; 
		\addlegendentry[]{\texttt{PLDM-0}} 
		
		\addplot[color = black, mark = triangle*, mark options = {scale = 1, fill = color1, solid}, line width = 0.3pt, smooth, densely dotted] table {data/berAverageSystemUnicastProposed1.txt}; 
		\addlegendentry[]{\texttt{PLDM-1}} 
		
		\addplot[color = black, mark = *, mark options = {scale = 0.8, fill = yellow, solid}, line width = 0.3pt, smooth, densely dotted] table {data/berAverageSystemUnicastProposed2.txt};
		\addlegendentry[]{\texttt{PLDM-2}} 
		
		\end{semilogyaxis}
		\end{tikzpicture}
		\centering
		\caption{Unicast}
	\end{subfigure}
	\caption{Bit Error Rate Performance}
	\label{f5}
\end{figure*}

Throughout the simulations, we consider the geometric channel model with $L = 8$ propagation paths (to depict the highly reflective industrial environment \cite{b5}), where the azimuth angles of departure and arrival are uniformly distributed over $\left[-\pi; \pi \right] $. Also, $ N_\mathrm{tx} = 64 $, $ L_\mathrm{tx} = 32 $, $ N_\mathrm{rx} = 4 $, $ L_\mathrm{rx} = 4 $ and $ K = 6 $. The maximum transmit power, the power splitting parameter, and the multicast SINR target are $ P_\mathrm{tx} = 1 \small{\text{W}} $, $ \beta = 3 $, $ \gamma_\mathrm{min} = 5 \small{\text{dB}}$, respectively. 
We denote our two proposed schemes by \texttt{PLDM-1} and \texttt{PLDM-2} as mentioned in Section \ref{s1c}. In addition, we include the outcomes of \cite{b7}, which we denote by \texttt{PLDM-0}. The results depicting spectral efficiency (SE) performance have been averaged over $ 1000 $ simulations. Fig. \ref{f2} shows the aggregate SE of the system, which is the sum of the unicast and multicast components, considering all the receivers. In terms of aggregate SE, the three schemes perform similarly. This a consequence of employing highly optimized precoders, where each expends the same amount of power $ P_\mathrm{tx} $ that is distributed among the two signals in different proportions.

In Fig. \ref{f3}, the SE of the multicast and unicast signals is displayed. The multicast SE per device is expected to be $ \rho = \log_2 \left( 1 + 10^{\gamma_\mathrm{min} / 10} \right) = 2.057 $ bps/Hz. Further, when all devices are considered the \emph{aggregate multicast SE} should be $ \rho \times K = 12.344 $ bps/Hz. This target is more tightly achieved by \texttt{PLDM-2}. Notice that both \texttt{PLDM-1} and \texttt{PLDM-2} are capable of prioritizing the multicast signal over the unicast counterpart so as to satisfy (\ref{e5b}). Nevertheless, \texttt{PLDM-1} can provide higher multicast SE than \texttt{PLDM-2} for low $ P_\mathrm{tx}/{\sigma^2} $. As $ P_\mathrm{tx}/{\sigma^2} $ increases, additional usable power becomes available for both signals to boost the SE. Thus, as the multicast signal approaches its target $ \gamma_\mathrm{min} $, it is progressively induced to a state where the SE saturates and the unicast SE gains more relevance. This behavior is attained through (\ref{e5a}), (\ref{e5b}) and (\ref{e5j}). On the contrary, in \texttt{PLDM-0} the unicast SE is higher than the multicast SE for low $ P_\mathrm{tx}/{\sigma^2} $, which is unsuitable. This undesirable behavior on data prioritization is obtained even though a weighted max-min approach was considered in \cite{b12}, aiming to emphasize the multicast importance.

In order to assess multicast fairness, Fig. \ref{f4} shows the SE for all devices within a confidence interval of $ 95\% $, where the shaded area delimitates the upper and lower bounds. We observe that \texttt{PLDM-1} outperforms \texttt{PLDM-2} at prioritizing multicast information for low $ P_\mathrm{tx}/{\sigma^2} $. Notice that once the target is reached, the SE of \texttt{PLDM-1} remains in the upper region of the desired threshold with some variability. On the other hand, \texttt{PLDM-2} is capable of maintaining a high multicast SE fairness among all the devices with negligible variance. For the sake of comparison, the results of \texttt{PLDM-0} are included.

Fig. \ref{f5} shows the bit error rate (BER) performance averaged over 10\textsuperscript{6} simulations, where the unicast and multicast symbols were obtained from a 4-QAM constellation. Since the multicast prioritization mechanism proposed in \cite{b12} does not work as expected, SIC cannot operate properly thus severely degrading the BER. We swapped (where necessary) the decoding order between unicast and multicast information to favor \texttt{PLDM-0}. On the other hand, \texttt{PLDM-1} and \texttt{PLDM-2} perform very similarly with a slight advantage of the former.

%% file: text/discussion.tex
\section{Discussion} \label{s6}

\noindent{\textbf{Unicast/multicast dichotomy:}}
Although the power splitting mechanism promotes the prioritization of the multicast signal, the multicast SINR and SE are not always higher than that of the unicast signal. This is advantageous since unicast transmissions can support higher order modulation in high SNR regime.

\noindent{\textbf{Multicast SINR threshold:}}
Having a deterministic $ \gamma_{\min} $ is justified since beacon control packets are usually of fixed size and a target SINR that allows successful decoding of the message can be derived. 

\noindent{\textbf{\texttt{PLDM-1} vs \texttt{PLDM-2}:}}
Recall that in \texttt{PLDM-1}, $ \mathbf{x} = \mathbf{F} \left[ \mathbf{B} \vert \mathbf{m} \right] \left[ \mathbf{s} \vert z \right]^T  = \mathbf{F} \mathbf{B} \mathbf{s} + \mathbf{F} \mathbf{m} z $. In \texttt{PLDM-2}, if $ \mathbf{m} = \mathbf{B}  \mathbf{u} $ then $ \mathbf{x} = \mathbf{F} \mathbf{B} \mathbf{s} + \mathbf{F} \mathbf{B} \mathbf{u} z = \mathbf{F} \mathbf{B} \left[ \mathbf{s} + \mathbf{u} z \right] $, where $ \mathbf{u} = \left[ u_1, \cdots, u_K \right]^T \succcurlyeq 0 $. Realize that $ u_k $ defines the ratio of energy between the unicast symbol $ s_k $ and the multicast symbol $ z $, for the $ k $-th device. Thus, \texttt{PLDM-2} has only one set of nearly-orthogonal digital unicast precoding vectors $ \left\lbrace \mathbf{b}_k \right\rbrace^K_{k = 1}  $ that are matched to the channel of each device. As a result, the multicast packet and the $ k $-th unicast information are conveyed simultaneously through the $ k $-th precoding vector $ \mathbf{b}_k $ with different powers (since $ \mathbf{m} = \mathbf{B}  \mathbf{u} $). On the contrary, \texttt{PLDM-1} is devised as non-orthogonally overlaid unicast and multicast precoders. Therefore, the spatial radiation patterns of  $ \left\lbrace \mathbf{b}_k \right\rbrace^K_{k = 1}  $ and $ \mathbf{m} $ do not necessary match.

\noindent{\textbf{\texttt{PLDM-0}:}}
In \cite{b12}, the authors attempted to prioritize multicast information by assigning (in the objective function) a higher weighting factor to the multicast minimum SINR. However, the formulation proposed therein did not allow to enforce such condition as observed in the simulations results.

\noindent{\textbf{Initialization:}} In order to solve $ \widetilde{\mathcal{P}}^{(t)}_2 $ we need an initial feasible point for $ \left\lbrace p^{(0)}_k \right\rbrace^K_{k = 1} $ and $ \mathbf{m}^{(0)} $. In this paper we selected, $ p^{(0)}_k = \frac{P_\mathrm{tx}}{2 \sum_{k} \left\| \mathbf{F} \mathbf{v}_k \right\|^2_2} $, $ \forall k \in \mathcal{K }$ and a randomly generated $ \mathbf{m} $ such that $ \left\| \mathbf{F} \mathbf{m}^{(0)} \right\|^2_2 = \frac{P_\mathrm{tx}}{2} $.

%% file: text/conclusion.tex
\section{Conclusion} \label{s7}
In this paper, we investigated the joint optimization of hybrid precoding, fairness, and power splitting in NOMA-LDM superimposed transmissions for industrial IoT scenarios. We proposed two solutions: one of them regarded as the superposition of two distinct precoders with different spatial and power signatures. The second approach is designed as a purely power-domain NOMA scheme. We included a power sharing constraint to support the SIC decoder task at the receiver. In addition, simulations show that both proposed schemes are capable of providing a high degree of fairness among all the devices, which is relevant for the dissemination of critical control messages in this kind of scenarios.

%% file: text/acknowledgment.tex
\section{Acknowledgment}
This research was in part funded by the Deutsche Forschungsgemeinschaft (DFG) within the B5G-Cell project as part of the SFB 1053 MAKI.

%% file: text/appendix.tex
\appendix
Let us define the function $ g : \mathcal{Z} \rightarrow \mathbb{R}^{+}_0 $ that maps any 4-tuple $ \left( \mathbf{w}_k, \mathbf{H}_k, \mathbf{F}, \mathbf{b}_k \right) \in \mathcal{Z} $ to $ \gamma_k $, where $ \mathcal{Z} $ is a subspace of $ \mathbb{C}^{N_\mathrm{rx} \times 1} \times \mathbb{C}^{N_\mathrm{rx} \times N_\mathrm{tx}} \times \mathbb{C}^{N_\mathrm{tx} \times K} \times \mathbb{C}^{K \times 1} $.

\emph{Lemma:} It holds true that the 4-tuple $ \left( \mathbf{w}_k, \mathbf{H}_k, \mathbf{F}, \mathbf{b}_k \right) \in \mathcal{Z} $ for which $\log_2 \left(  g \left( \mathbf{w}_k, \mathbf{H}_k, \mathbf{F}, \mathbf{b}_k \right) \right) $ is maximal in $ \mathcal{Z} $, also makes $ g \left( \mathbf{w}_k, \mathbf{H}_k, \mathbf{F}, \mathbf{b}_k \right) $ maximal in $ \mathcal{Z} $.

\emph{Proof:} Since every $ g \left( \mathbf{w}_k, \mathbf{H}_k, \mathbf{F}, \mathbf{b}_k \right) \in \mathbb{R}^{+}_0 $, then $ g \left( \mathbf{w}_k, \mathbf{H}_k, \mathbf{F}, \mathbf{b}_k \right) \geq g \left( \widehat{\mathbf{w}}_k, \widehat{\mathbf{H}}_k, \widehat{\mathbf{F}}, \widehat{\mathbf{b}}_k \right) \Leftrightarrow \log_2 \left( g \left( \mathbf{w}_k, \mathbf{H}_k, \mathbf{F}, \mathbf{b}_k \right) \right) \geq \log_2 \left( g \left( \widehat{\mathbf{w}}_k, \widehat{\mathbf{H}}_k, \widehat{\mathbf{F}}, \widehat{\mathbf{b}}_k \right) \right) $, due to $ \log_2 \left( \cdot \right) $ being monotonically increasing in $ \mathbb{R}^{+}_0 $. Thus, for every  $ \left( \widehat{\mathbf{w}}_k, \widehat{\mathbf{H}}_k, \widehat{\mathbf{F}}, \widehat{\mathbf{b}}_k \right) \in \mathcal{Z} $, $ g \left( \mathbf{w}_k, \mathbf{H}_k, \mathbf{F}, \mathbf{b}_k \right) \geq g \left( \widehat{\mathbf{w}}_k, \widehat{\mathbf{H}}_k, \widehat{\mathbf{F}}, \widehat{\mathbf{b}}_k \right) $. Also, for every $ \left( \widehat{\mathbf{w}}_k, \widehat{\mathbf{H}}_k, \widehat{\mathbf{F}}, \widehat{\mathbf{b}}_k \right) \in \mathcal{Z}, \log_2 \left( g \left( \mathbf{w}_k, \mathbf{H}_k, \mathbf{F}, \mathbf{b}_k \right) \right) \geq \log_2 \left( g \left( \widehat{\mathbf{w}}_k, \widehat{\mathbf{H}}_k, \widehat{\mathbf{F}}, \widehat{\mathbf{b}}_k \right) \right) $. Therefore, the 4-tuple $ \left( \mathbf{w}_k, \mathbf{H}_k, \mathbf{F}, \mathbf{b}_k \right) \in \mathcal{Z} $ that maximizes $\log_2 \left(  g \left( \mathbf{w}_k, \mathbf{H}_k, \mathbf{F}, \mathbf{b}_k \right) \right) $, also maximizes $ g \left( \mathbf{w}_k, \mathbf{H}_k, \mathbf{F}, \mathbf{b}_k \right) $ and these two inequalities are equivalent.